%% file: Kalhor_Thundat_Jacob_arXiv.tex
\documentclass[%
 aip,
 amsmath,amssymb,
 reprint,%
]{revtex4-1}
\usepackage{graphicx}
\usepackage{dcolumn}
\usepackage{bm}
\usepackage{epstopdf}

\begin{document}

\preprint{AIP/123-QED}


\title[Universal Spin-Momentum Locked Optical Forces]{Universal Spin-Momentum Locked Optical Forces}



\author{Farid Kalhor}
\affiliation{Department of Electrical and Computer Engineering, University of Alberta, Edmonton, Alberta T6G 1H9, Canada}

\author{Thomas Thundat}
\affiliation{Department of Chemical and Materials Engineering, University of Alberta, Edmonton, Alberta T6G 1H9, Canada}

\author{Zubin Jacob}
\thanks{Corresponding author: zjacob@purdue.edu}
\affiliation{Birck Nanotechnology Center, Department of Electrical and Computer Engineering, Purdue University, West Lafayette, Indiana 47906, USA}
\affiliation{Department of Electrical and Computer Engineering, University of Alberta, Edmonton, Alberta T6G 1H9, Canada}



\begin{abstract}

Evanescent electromagnetic waves possess spin-momentum locking, where the direction of propagation (momentum) is locked to the inherent polarization of the wave (transverse spin). We study the optical forces arising from this universal phenomenon and show that the fundamental origin of recently reported non-trivial optical chiral forces is spin-momentum locking. For evanescent waves, we show that the  direction of energy flow, direction of decay, and direction of spin follow a right hand rule for three different cases of total internal reflection, surface plasmon polaritons, and $HE_{11}$ mode of an optical fiber.
Furthermore, we explain how the recently reported phenomena of lateral optical force on chiral and achiral particles is caused by the transverse spin of the evanescent field and the spin-momentum locking phenomenon. Finally, we propose an experiment to identify the unique lateral forces arising from the transverse spin in the optical fiber and point to fundamental differences of the spin density from the well-known orbital angular momentum of light. Our work presents a unified view on spin-momentum locking and how it affects optical forces on chiral and achiral particles.
  
\end{abstract}


\maketitle 

Evanescent electromagnetic waves possess a unique intrinsic property of spin-momentum locking where the direction of phase propagation, which can be associated with momentum, is locked to the direction of an intrinsic polarization \cite{bliokh_extraordinary_2014, bliokh_transverse_2012, van_mechelen_universal_2015, kien_negative_2013}. The similarities with surface states of electronic topological insulators led to the coining of the term  - Quantum spin-hall effect of light \cite{bliokh_quantum_2015, van_mechelen_universal_2015, aiello_transverse_2009}. This intrinsic polarization of evanescent waves can be associated with a spin vector which is transverse to both the direction of its momentum and the direction of decay \cite{bliokh_quantum_2015, van_mechelen_universal_2015}. The evanescent waves also transfer the spin-momentum locking to bulk modes through electromagnetic boundary conditions \cite{van_mechelen_universal_2015}. The phenomenon of spin-momentum locking of light has been observed in a variety of systems including optical fibers, surface plasmon-polaritons and photonic crystal waveguides\cite{rodriguez-fortuno_near-field_2013, kapitanova_photonic_2014, mitsch_quantum_2014, petersen_chiral_2014, le_feber_nanophotonic_2015, sollner_deterministic_2015, young_polarization_2015, aiello_transverse_2015}. A unifying theme in these experiments is that scattered or spontaneously emitted light with an intrinsic polarization shows unidirectional propagation.

It is an interesting question whether this intrinsic polarization of the evanescent wave (transverse spin) can lead to optical forces and whether they show fundamental behavior different from well-known gradient and radiation pressure forces \cite{ashkin_acceleration_1970, ashkin_observation_1986}. Very recent studies involving optical forces on chiral particles, which contain both induced electric and magnetic moments, show that these particles can experience an optical force perpendicular to the direction of the momentum of the incident field \cite{wang_lateral_2014, hayat_lateral_2015,cameron_discriminatory_2014}. It was recently shown that the particle in the unidirectional coupling experiment also experiences an anomalous optical force perpendicular to the direction of the momentum of the incident beam \cite{rodriguez-fortuno_lateral_2015}. These anomalous forces cannot be explained by the gradient or radiation pressure forces.

In this paper, we present a unified view of anomalous optical forces on chiral and achiral particles and uncover the fundamental connection to the spin-momentum locking phenomenon associated with evanescent waves. This force is different in nature from the gradient force or the optical pressure and is in fact related to the transverse spin of evanescent  electromagnetic fields interacting with the particle. We make the important observation that there exists a fundamental right handed triplet vector for evanescent waves related to momentum, decay and transverse spin which governs the direction of optical forces.  The spin-density forces are rigorously calculated using a stress tensor formalism for evanescent waves generated during total internal reflection (TIR) and the near-field of an optical fiber. Subtle connections between spin-momentum locking and the non-trivial direction of the spin forces are revealed paving the way for experimental demonstrations. Our work also shows clearly the independence of the orbital angular momentum (OAM) of the propagating optical fiber mode and the transverse spin-density of the evanescent waves associated with the near-field.


We start with a discussion of the field quantities defining the intrinsic spin of electromagnetic evanescent waves. This is a subtle question and can only be answered by making connections to experimental observables such as energy and optical forces. Momentum density of electromagnetic waves is proportional to the Poynting vector \cite{aiello_transverse_2009}, $\langle\vec{p}\rangle=\langle\vec{N}\rangle/c^2$, where $\langle\vec{p}\rangle$ is the time averaged momentum density, $\langle\vec{N}\rangle=1/2Re\big[\vec{E} \times \vec{H}^*\big]$ is the time averaged Poynting vector, and $c$ is the speed of light. For evanescent waves, the direction of this momentum is always perpendicular to the direction of decay and is the same as the direction of the real part of the wave vector \cite{van_mechelen_universal_2015}. The time averaged spin density of evanescent waves has been defined by \cite{berry_optical_2009, bliokh_extraordinary_2014, neugebauer_measuring_2015}, $\langle\vec{s}\rangle=\left(1/2\omega \right)Im\big[\epsilon_0 \vec{E}^*\times \vec{E}+\mu_0 \vec{H}^*\times \vec{H}\big]=\langle\vec{s}_e\rangle+\langle\vec{s}_m\rangle$, where $\langle \vec{s}_e\rangle$ and $\langle \vec{s}_m\rangle$ are the electric and magnetic spin densities, respectively.
Another equivalent approach is to define the electric and magnetic spin densities by introducing the generalized electric and magnetic stokes parameters \cite{_see_????, van_mechelen_universal_2015, setala_degree_2002, bekshaev_transverse_2007}, $\langle \vec{s}_e\rangle=\vec{S}^e_3 /(\omega c)$ and $\langle \vec{s}_m\rangle=\vec{S}^m_3 /(\omega c)$. 

\begin{figure}[htbp]
\centering
  \begin{tabular}{cc}

    \includegraphics[width=0.6\linewidth, trim={0 0 0 0},clip]{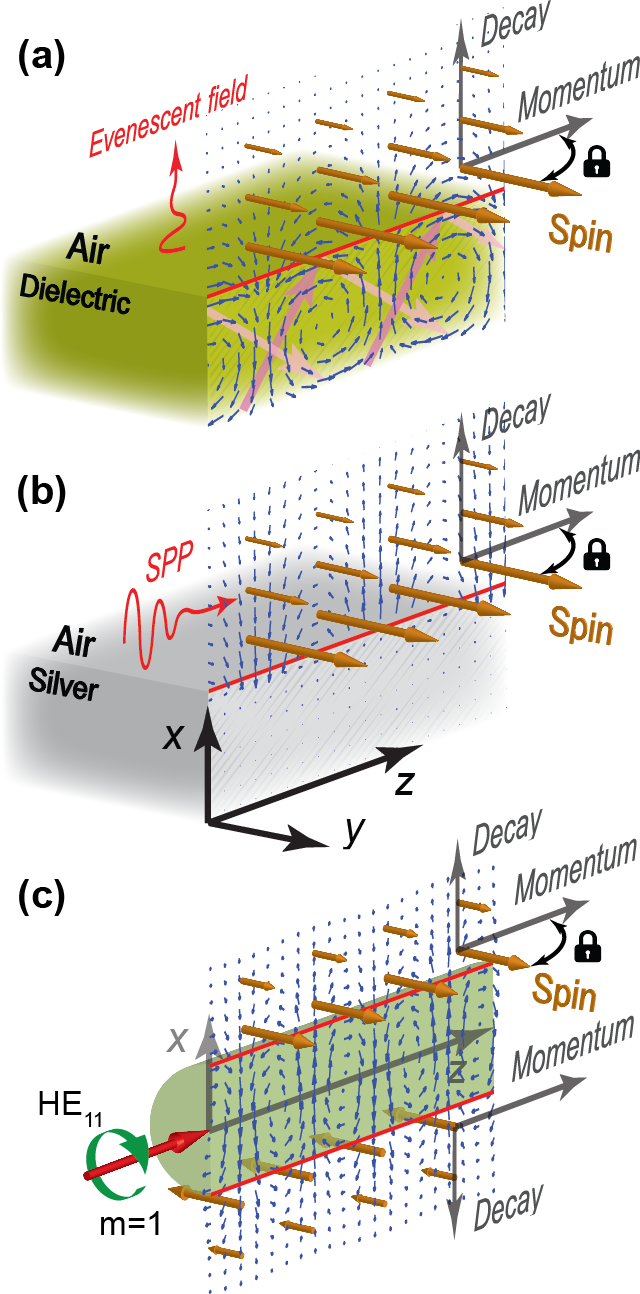}

  \end{tabular}
\caption{Direction of transverse spin in evanescent fields for (a) total internal reflection, (b) surface plasmon polariton, and (c) optical fiber. The blue vectors show the electric field at an instance of time and the orange vectors show the direction of transverse spin. In all cases the direction of momentum, direction of decay, and transverse spin follow a right hand rule.}
\label{fig:evanescent_spin}
\end{figure}

In Figure \ref{fig:evanescent_spin} the direction of the component of the spin density transverse to the plane of momentum and decay direction is shown for three different cases. This figure illustrates that the triplet of momentum, decay direction and transverse spin follows a right-hand rule. This means for a beam propagating in $+\hat{z}$ direction if the wave is decaying in $+\hat{x}$ direction the transverse spin will be in $+\hat{y}$ direction. We emphasize the subtle observation that the opposite direction of the transverse spin  is inconsistent with causality because it refers to a wave growing in magnitude away from the boundaries.

Spin-momentum locking has profound consequences on optical forces. Optical force acting on a chiral particle in an incident electromagnetic field, with induced dipole moments $\begin{bmatrix} \vec{p} \\ \vec{m} \end{bmatrix} = \begin{bmatrix} \alpha_{ee} & i\alpha_{em} \\ -i\alpha_{em} & \alpha_{mm} \end{bmatrix} \begin{bmatrix} \vec{E} \\ \vec{H} \end{bmatrix}$, can be expressed as \cite{wang_lateral_2014, _see_????} $\langle\vec{F} \rangle=\langle\vec{F}_{gr} \rangle+\langle\vec{F}_{op} \rangle+\langle\vec{F}_{sr} \rangle$
where all the forces are time averaged. 
$\langle\vec{F}_{gr} \rangle$ and $\langle\vec{F}_{op} \rangle$ are the gradient force and radiation pressure, respectively.
$\langle \vec{F}_{sr} \rangle=-(c k_0^4/6\pi) \Big\{ \big( Re[\alpha_{ee} \alpha_{mm}^*]+|\alpha_{em}|^2 \big) \langle \vec{N} \rangle + \sqrt{\mu_0/\epsilon_0} Re[\alpha_{ee}\alpha_{em}^*] \vec{S}^e_3 + \sqrt{\epsilon_0/\mu_0} Re[\alpha_{mm}\alpha_{em}^*] \vec{S}^m_3 -1/2Im[\alpha_{ee}\alpha_{mm}^*]Im[\vec{E}\times\vec{H}^*] \Big\}$ 
is the scattering recoil force.
It vanishes for an achiral particle with no magnetic polarizability. In the case of a chiral particle, this force has two terms proportional to the electric and magnetic fourth Stokes parameters (or equivalently electric and magnetic spin densities). The sum of these two terms is known as the spin density force. Our aim is to show that the subtle phenomenon of spin-momentum locking dictates the non-trivial direction of this force in various scenarios where nanoparticles are interacting with evanescent waves.

\begin{figure}[htbp]
\centering
  \begin{tabular}{cc}

    \includegraphics[width=0.6\linewidth]{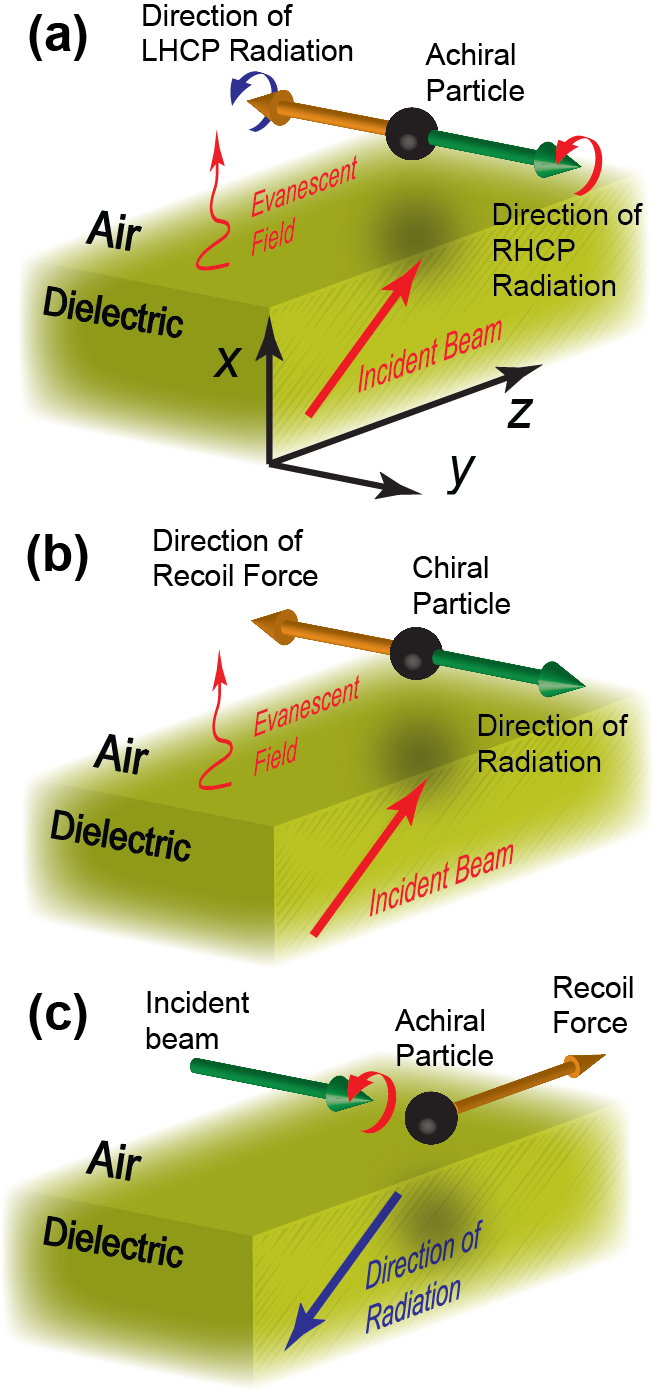}

  \end{tabular}

\caption{Lateral force caused by transverse spin of the evanescent fields.
(a) An achiral particle is placed in an evanescent field. The scattered radiation by the particle has a circular polarization. The handedness of the scattered radiation follows the direction of spin and is locked to the direction of momentum of the incident field.
(b) A chiral particle is placed in an evanescent field experiences a lateral force. The lateral force is proportional to the transverse spin of the evanescent field (spin in $\hat{y}$ direction). The spin of the evanescent field causes a directional scattering by the particle. The direction of the scattered radiation is parallel (or anti-parallel depending on the handedness of the chiral particle) to the direction of spin, therefore, the recoil force caused by the directional radiation will be anti-parallel (parallel) to the direction of spin.
(c) Reciprocal setup of (a). An achiral particle placed near a dielectric surface is illuminated by a circularly polarized beam. The scattered radiation by the particle propagates directionally inside the dielectric. The direction of propagation, with respect to $\hat{z}$ only depends on the polarization of the incident beam. The particle experiences a recoil force in the opposite direction of propagation of the scattered field.}
\label{fig:force_schem}
\end{figure}

We first consider the phenomenon of total internal reflection at a dielectric-air interface with a nanoparticle placed on the air side as shown in Figure \ref{fig:force_schem}. The evanescent waves generated  have the same direction of propagation parallel to the interface as the incident beam i.e. their momentum is fixed. This immediately fixes the transverse spin direction of the evanescent wave by the spin-momentum locking phenomenon. Consequently, light scattered by chiral and achiral particles will show fundamentally different behavior since the incident evanescent wave is intrinsically handed. In Figure \ref{fig:force_schem}(a), the achiral particle scatters the evanescent wave equally in two directions leading to zero net momentum in direction transverse to the plane of incidence and therefore no recoil force in this direction. However, Figure \ref{fig:force_schem}(b) shows the case of a chiral particle which interacts with the spin of the evanescent waves and scatters the light asymmetrically leading to a ``sideways" (lateral)  recoil force i.e. transverse to the incident plane. This phenomenon is the cause of the anomalous lateral force reported in recent works \cite{wang_lateral_2014, hayat_lateral_2015}. The resulting force will be in opposite directions for the two enantiomers of a chiral particle.  To obtain a lateral force on an achiral particle, we can use the reciprocal experiment of Figure \ref{fig:force_schem}(a) shown in Figure \ref{fig:force_schem}(c). This means, if we illuminate an achiral particle placed near an air/dielectric interface with right-handed circularly polarized light propagating in $+\hat{y}$ direction, the scattered evanescent fields will be right-handed i.e. their spin is fixed. This fixes the direction of momentum parallel to the interface due to spin-momentum locking and the tunneled evanescent waves propagate preferentially with a net momentum in $-\hat{z}$ direction. This directional scattering will apply a recoil force on the particle in $+\hat{z}$ direction. Note the right-hand rule from Figure \ref{fig:evanescent_spin} is central to this interplay of spin and momentum of incident light and lateral recoil force.

\begin{figure}[htbp]
\centering
  \begin{tabular}{cc}

    \includegraphics[width=\linewidth]{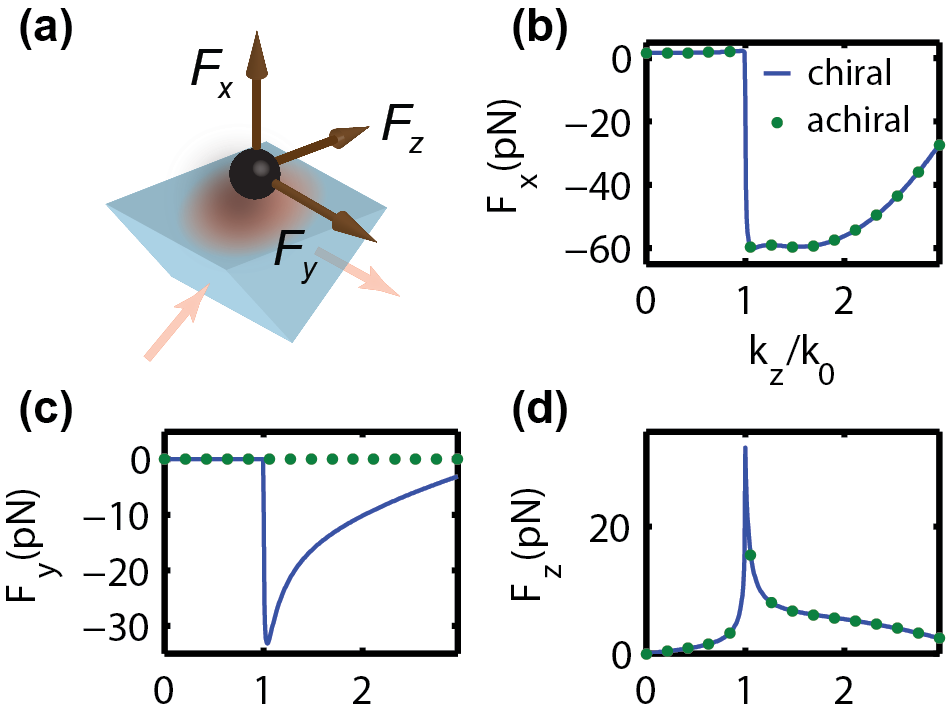}

  \end{tabular}
\caption{Optical forces on a chiral particle (solid line) and an achiral particle (dots) in an evanescent field. (a) Schematic of a chiral particle placed near a prism showing the direction of incident beam and the forces. (b) Time averaged force in $\hat{x}$ direction. This component of the force is dominated by the gradient force and is toward the prism. (c) Time averaged force in $\hat{y}$ direction. This component of the force is proportional to the transverse spin. (d) Time averaged force in $\hat{z}$ direction. This component of the force is dominated by radiation pressure. The particle is $100nm$ away from the surface of the prism. The polarizability matrix for the chiral particle is \cite{_see_????} $ \begin{bmatrix} \alpha_{ee} & i\alpha_{em} \\ -i\alpha_{me} & \alpha_{mm} \end{bmatrix} = \begin{bmatrix} 1.2 + i 0.1 & i0.01 \\ -i0.01 & 0.0002 \end{bmatrix} \times 10^{-24}$ and the polarizability of the achiral particle is $\alpha=(1.2+i0.1)\times10^{-24}$. The incident beam is $p$ polarized with wavelength $\lambda=5\mu m$ and carries a power of $p=50\ mW/mm^2$. The refractive index of the prism is $n=3.5$ and the horizontal axis is related to the incident angle of the beam.}
\label{fig:force_prism}
\end{figure}

The effect of the chirality of a particle on the transverse spin density force is illustrated in Figure \ref{fig:force_prism}, where the forces on a chiral and an achiral particle are compared.
By calculating the force in a purely evanescent wave we show that the origin of the aforementioned lateral optical force is the transverse spin of evanescent waves. Figure \ref{fig:force_prism}(c) shows that the lateral force only appears when the angle of incidence is larger than the critical angle of the prism/air interface. In this regime, the power is fully reflected and only evanescent fields exist beyond the prism surface. Furthermore, this figure shows that only a chiral particle can probe the transverse spin of an evanescent wave if the incident wave is linearly polarized.

\begin{figure}[htbp]
\centering
  \begin{tabular}{cc}
    \includegraphics[width=\linewidth]{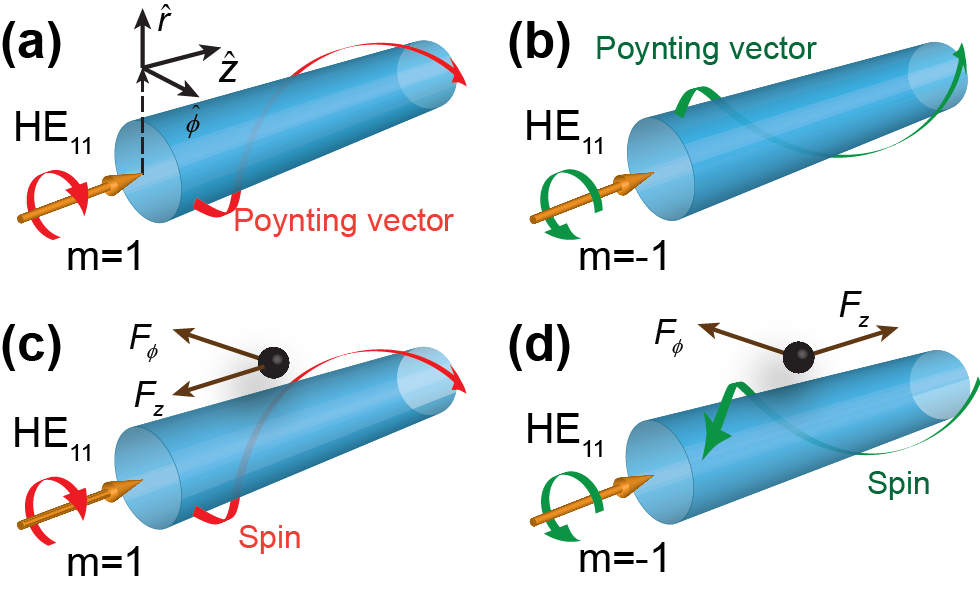}
  \end{tabular}
\caption{Direction of Poynting vector, spin density and spin density force on a chiral particle for $m=1$ and $m=-1$ $HE_{11}$  modes. (a) and (b) Direction of Poynting vector outside the optical fiber for the two modes. (c) and (d) Direction of spin density and spin density force outside the optical fiber for the two modes. The longitudinal component of the spin density has opposite directions for the two modes, but, its transverse component is in the same direction for both modes which shows the spin-momentum locking in evanescent fields. The direction of spin density force on a chiral particle placed outside the optical fiber is also shown. The longitudinal component of the force, $F_z$, has opposite directions for the two modes, but, the transverse component, $F_\phi$, is in the same direction for both modes.}
\label{fig:fiber_schem}
\end{figure}

We now show effects of spin-momentum locking phenomenon on optical forces  in the cylindrically symmetric case of an optical fiber. The fundamental mode of an optical fiber is the $HE_{11}$ mode \cite{agrawal_fiber-optic_2002} and can appear in two degenerate forms of $m=1$ and $m=-1$ where $m$ is related to the OAM of the mode. Figure \ref{fig:fiber_schem} depicts how this OAM related to the propagating wave field is fundamentally different from the transverse spin density related to the evanescent waves present outside the core of the fiber. This is revealed through the arrows depicting the contrasting nature of energy flow and spin-density flow when the OAM is reversed. The direction of the twist of the transverse spin-density around the fiber is fundamentally tied to the momentum and is unrelated to the OAM. A detailed analysis will be presented in future work and here we focus only on the consequences of this phenomenon on the non-trivial direction of optical forces on a chiral particle in the near-field of an optical fiber.

\begin{figure}[htbp]
\centering
  \begin{tabular}{cc}

    \includegraphics[width=\linewidth]{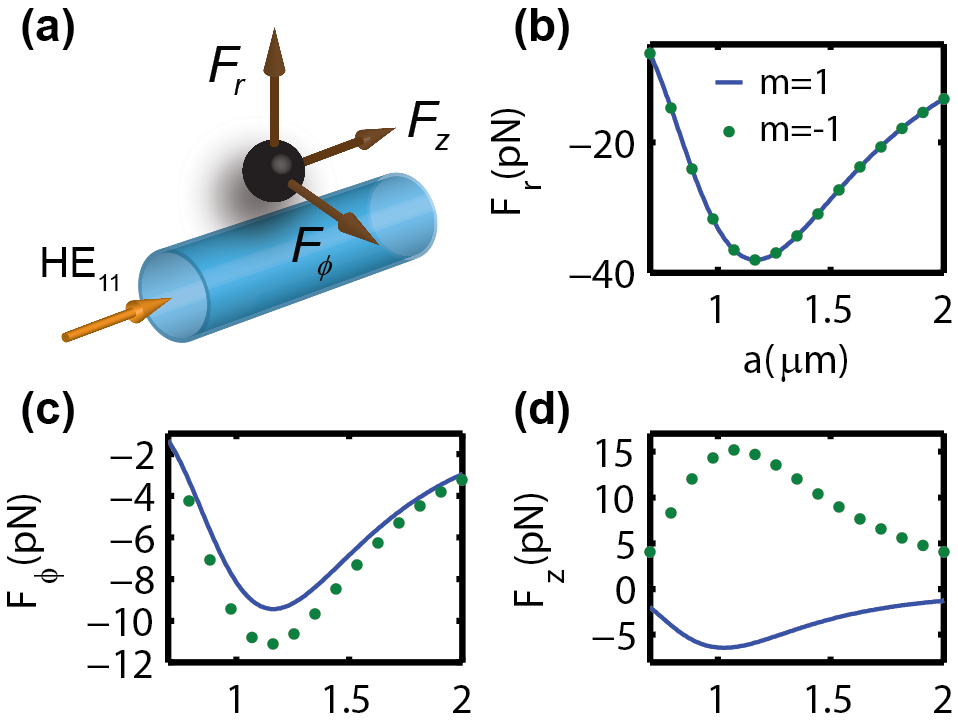}

  \end{tabular}
\caption{(a) Schematic of a chiral particle near an optical fiber. (b), (c) and (d) Time averaged optical force in different directions plotted versus radius of the fiber. The solid line shows the force for $m=1$ mode and the dots show the force for $m=-1$ mode. The $\hat{r}$ component is dominated by the gradient force and has the same value for both modes while the $\hat{\phi}$ component is dominated by the spin density force. The transverse component of the Poynting vector applies a radiation pressure force which is responsible for the difference in the $\hat{\phi}$ component of the force for the two modes. The direction of the transverse spin and transverse spin density force is independent of the OAM of the mode and is locked to its direction of propagation. Both radiation pressure and the spin density forces contribute to the $\hat{z}$ component of the force. The radiation pressure is in $+\hat{z}$ direction for both modes but the spin density force has opposite directions. Therefore, for the $m=1$ mode the force is in $+\hat{z}$ direction but for the $m=-1$ mode the force in the opposite direction. The particle is placed $100nm$ away from the surface of the fiber and has the same properties as the particle in Figure \ref{fig:force_prism}. The refractive index of the fiber is $n_1=1.5$ and the surrounding medium is vacuum. The wavelength of the incident beam is $\lambda=5\mu m$ and the mode carries a total power of $P=1 \mu W$.}
\label{fig:force_fiber}
\end{figure}

To compare the magnitude of the spin-momentum locked optical force with the other forces acting on a chiral particle, we have calculated the total force in different directions on a chiral particle placed near an optical fiber (Figure \ref{fig:force_fiber}). This figure compares the force for $m=1$ and $m=-1$ modes. The radial component of the force $F_r$ is dominated by the gradient force. Since the spin density force has no contribution in this component of the force, it has the same magnitude for both the modes.
In Figure \ref{fig:force_fiber}(d) we show the striking contrast between positive and negative angular momentum modes. The longitudinal force $F_z$ changes sign in accordance with the angular momentum mode. Its behavior is fundamentally different from the transverse spin density force.
Figure \ref{fig:force_fiber}(c) is the non-trivial lateral force which is related to the transverse spin-density. Its direction is independent of the angular momentum and is related only to the propagation direction in the fiber. The difference between the magnitude of this component for the two modes is because the $\hat{\phi}$ component of the radiation pressure has different directions for the two modes (Figure \ref{fig:fiber_schem}).

In conclusion, we have demonstrated universality of spin-momentum locking and the resulting lateral optical force on chiral particles in evanescent fields. We have shown that the phenomena giving rise to unidirectional coupling is also at the heart of the lateral optical force. The lateral optical force can be harnessed for manipulation of small particles and chirality sorting in evanescent wave fields. 

This work was supported by IC-IMPACTS. The authors want to thank A. Sihvola for useful informations and S. Jahani and C. Cortes for fruitful discussions.

\clearpage

\begin{widetext}

\input{supplimental_materials.tex}
\end{widetext}

\bibliographystyle{ieeetr}
\bibliography{transversespin}

\end{document}

%% file: supplimental_materials.tex
\LARGE
\noindent \textbf{Supplemental material note 1}

\bigskip
\Large
\noindent \textit{Electric and magnetic Stokes parameters}

\smallskip
\normalsize

We define the fourth electric and magnetic Stokes parameters as \cite{van_mechelen_universal_2015, setala_degree_2002, bekshaev_transverse_2007}:
$$\vec{S}_3^e=\frac{1}{2}\sqrt{\frac{\epsilon_0}{\mu_0}} Im\left( \vec{E}^*\times\vec{E} \right)=\omega c \langle\vec{s}_e \rangle ,$$
$$\vec{S}_3^m=\frac{1}{2}\sqrt{\frac{\mu_0}{\epsilon_0}} Im\left( \vec{H}^*\times\vec{H} \right)=\omega c \langle\vec{s}_m \rangle .$$
where the total fourth Stokes parameter is $\vec{S}_3=\vec{s}_e +\vec{s}_m$. This expression gives the same value as the conventional Stokes parameters for a plane wave.

\clearpage

\LARGE
\noindent \textbf{Supplemental material note 2}

\bigskip
\Large
\noindent \textit{Optical force on a chiral particle}

\medskip
\normalsize

The optical force on chiral particle given by Wang and Chan \cite{wang_lateral_2014} can be written in the form:
$$\langle\vec{F} \rangle=\langle\vec{F}_{gr} \rangle+\langle\vec{F}_{op} \rangle+\langle\vec{F}_{sr} \rangle$$
where $\langle \vec{F}_{gr} \rangle=\nabla U$ is the gradient force and
$$U=1/4\big(Re[\alpha_{ee}]|\vec{E}|^2+Re[\alpha_{mm}]|\vec{H}|^2-2Re[\alpha_{em}]Im[\vec{H} \cdot \vec{E}^*]\big).$$

$\langle \vec{F}_{op} \rangle$ is optical pressure force defined as:
\begin{equation*}
\begin{split}
\langle\vec{F}_{op} \rangle=\frac{k_0}{c} \left( \frac{Im[\alpha_{ee}]}{\epsilon_0} + \frac{Im[\alpha_{mm}]}{\mu_0}\right) \langle \vec{N}\rangle-Im[\alpha_{em}]\nabla\times \langle\vec{N}\rangle \\
 - \frac{c k_0}{2} \left( \frac{Im[\alpha_{ee}]}{\epsilon_0} \nabla\times \langle\vec{s}_e \rangle+ \frac{Im[\alpha_{mm}]}{\mu_0} \nabla\times \langle\vec{s}_m \rangle \right) \\
+ \omega^2 Im[\alpha_{em}] \left( \langle \vec{s}_e  \rangle + \langle \vec{s}_m \rangle \right)
\end{split}
\end{equation*}
where $k_0=\omega/c$.

$\langle \vec{F}_{sr} \rangle$ is scattering recoil force defined as:
\begin{equation*}
\begin{split}
\langle \vec{F}_{sr} \rangle=-\frac{c k_0^4}{6\pi} \Bigg\{ \left( Re[\alpha_{ee} \alpha_{mm}^*]+|\alpha_{em}|^2 \right) \langle \vec{N} \rangle + \sqrt{\frac{\mu_0}{\epsilon_0}} Re[\alpha_{ee}\alpha_{em}^*] \vec{S}^e_3 \\
 + \sqrt{\frac{\epsilon_0}{\mu_0}} Re[\alpha_{mm}\alpha_{em}^*] \vec{S}^m_3 -\frac{1}{2}Im[\alpha_{ee}\alpha_{mm}^*]Im[\vec{E}\times\vec{H}^*] \Bigg\}.
\end{split}
\end{equation*}

\clearpage

\LARGE
\noindent \textbf{Supplemental material note 3}

\bigskip
\Large
\noindent \textit{Polarizability of particle}

\medskip
\normalsize

The polarizability of the particle is an approximate value for a spherical particle of a volume $V=1000nm^3$. For such a particle the polarizability matrix is:
$$ \begin{bmatrix} \alpha_{ee} & i\alpha_{em} \\ -i\alpha_{me} & \alpha_{mm} \end{bmatrix} = \begin{bmatrix} 1.2 + i 0.1 & i0.01 \\ -i0.01 & 0.0002 \end{bmatrix} \times 10^{-24}$$
The relative permittivity of the particle is assumed to be $\epsilon_r=3+i0.28$ and the electric-electric polarizability factor $\alpha_{ee}$ is calculated using \cite{maier_plasmonics:_2007} $\alpha_{ee}=3V (\epsilon_r-\epsilon_m)/(\epsilon_r+2\epsilon_m)$, where $\epsilon_m=1$ is the relative permittivity of the vacuum.

The chirality factor $\alpha_{em}$ is calculated using the parameters of Hexahelicene molecule \cite{barron_molecular_2004, cameron_discriminatory_2014} at wavelength $\lambda=5 \mu m$. The density of the particle is approximated with that of water.

The magnetic polarizability $\alpha_{mm}$ is chosen such to satisfy the condition $\alpha_{ee} \alpha_{mm} \geq \alpha_{em} \alpha_{me}$. For a helical particle \cite{lindell_electromagnetic_1994} $\alpha_{ee} \alpha_{mm} = \alpha_{em} \alpha_{me}$.

\clearpage

